# HIGH-DIMENSIONAL ITERATIVE VARIABLE SELECTION FOR ACCELERATED FAILURE TIME MODELS

[1]NILOTPAL SANYAL

[1]Vitising Scientist, Indian Statistical Institute, 203 B. T. Road, Kolkata 700037, India

Email: [1]nilotpal.sanyal@gmail.com
Contact: [1](+91) 629-054-9681

**Abstract:** We propose an iterative variable selection method for the accelerated failure time model using high-dimensional survival data. Our method pioneers the use of the recently proposed structured screen-and-select framework for survival analysis. We use the marginal utility as the measure of association to inform the structured screening process. For the selection steps, we use Bayesian model selection based on non-local priors. We compare the proposed method with a few well-known methods. Assessment in terms of true positive rate and false discovery rate shows the usefulness of our method. We have implemented the method within the R package GWASinlps.

## I. INTRODUCTION

We consider the problem of variable selection in survival data where the outcome variable is time to some event, often called 'failure' (e.g., biological death in a clinical trial). Many clinical studies that contain survival information for the event of interest may have access to complementary subject-specific information from imaging data or public health data on a large number of covariates some of which may be associated with the survival time. The covariates may, for example, be SNP genotypes from genomics data, gene expressions from microarray data, or clinical, behavioral and historical variables from healthcare data. Several frequentist and Bayesian methods are available for high-dimensional variable selection in survival data (see the references within [1]). Recently, a high-dimensional variable selection method for continuous outcomes is proposed by [2] in the context of genome-wide association studies and further extended to binary outcomes by [3]. They introduce the concept of a 'structured screen-and-select' strategy and examine the use of non-local priors within the same. In this work, we extend their method to the analysis of survival data by proposing an iterative variable selection method based on non-local priors for accelerated failure time models.

## II. METHODS

Suppose we have $n$ subjects and for each subject, information on $p$ covariates where p can be much larger than n (p ≫ n). Suppose $x_i$ is the vector of covariates for subject $i$, $i = 1, ..., p$, and $X = (x_1, ..., x_p)$. Let $t$ denote the failure time and $c$ denote the censoring time; they are subscripted by $i$ for the $i$th subject.

Each iteration of the proposed method comprises two parts—(i) the *screening part* that screens a smaller number of variables from all available candidate variables based on some measures of association, and (ii) the *selection part* that selects variables from the screened variables based on non-local prior-based Bayesian model selection [4]. Below, we describe the accelerated failure time model and the non-local priors used for the selection part, and subsequently, the proposed method.

## II. A. ACCELERATED FAILURE TIME MODELS

Let $p_s$ generically denote the number of variables entering a selection step. For the variable selection context, let $k$ index a model where $k = 1, \ldots, 2^{p_s}$ and let $n_k$ denote the number of variables in model $k$. For any $k$, we consider an accelerated failure time (AFT) model with log-normal distribution for the failure time $t$. The AFT model can be written as a linear model for the logarithm of $t$, given by

$$\log(t_i) = \mu + \boldsymbol{x}_{ik}^T \boldsymbol{\beta}_k + \sigma Z,$$

where $Z \sim N(0,1)$, $\boldsymbol{\beta}_k = (\beta_{1,k}, \ldots, \beta_{n_k,k})$ is the vector of regression parameters for the covariates of model $k$, and $\mu$ and $\sigma$ are respectively an intercept and a scale parameter. The survival function for the above AFT model is given by

$$S(t_i) = 1 - \Phi\left\{\frac{\log(t_i) - (\mu + \boldsymbol{x}_{ik}^T \boldsymbol{\beta}_k)}{\sigma}\right\},$$

where $\Phi$ is the cumulative distribution function of the standard normal distribution.

Suppose $(y_i, x_i, \delta_i)$, $i = 1, \ldots, n$, are the observed data where $y_i = \min(t_i, c_i)$ and $\delta_i = 0\ or\ 1$ according as whether the $i$th subject is censored ($t_i > c_i$) or not. The likelihood of the above AFT model is given by

$$L(\beta) = \prod_{i=1}^{n} \{f(y_i)^{\delta_i} S(y_i)^{1-\delta_i}\},$$

where $f(.)$ is the probability density function of the log-normal distribution. The log-likelihood can be written as (ignoring the terms independent of $\beta$)

$$l(\beta) = \sum_{i=1}^{n} \left[ -\frac{\delta_i}{2\sigma^2} \{\log(y_i) - (\mu + \boldsymbol{x}_{ik}^T \boldsymbol{\beta}_k)\}^2 \right.$$
$$\left. - (1 - \delta_i) \log\left(1 - \Phi\left\{\frac{\log(y_i) - (\mu + \boldsymbol{x}_{ik}^T \boldsymbol{\beta}_k)}{\sigma}\right\}\right) \right].$$

## II. B. NON-LOCAL PRIORS

In the Bayesian model selection, prior distributions need to be specified over both the model space and the parameter space. Here, we consider three different non-local priors over the parameter space—the product moment prior (pMOM prior), the product inverse moment prior (piMOM prior), and the product exponential moment prior (peMOM prior) [4]. The pMOM prior, which is the product of individual moment (MOM) priors for the regression parameters $\boldsymbol{\beta}_k$, is given by

$$\pi_M(\boldsymbol{\beta}_k|r,\tau_1) = M^{-1}(\phi\tau_1)^{-\frac{n_k}{2}-rn_k}\prod_{j=1}^{n_k}\beta_{j,k}^{2r}\exp\left[-\frac{1}{2\phi\tau_1}\sum_{j=1}^{n_k}\beta_{j,k}^2\right],$$

where $\tau_1$ is the common scale parameter and $r$ is the common order of the MOM priors, $\phi$ is a dispersion parameter and $M$ is a marginalizing constant given by $M = (2\pi)^{-n_k/2}\prod_{l=1}^{r}(2l-1)^{n_k}$.

The piMOM prior, which is the product of individual inverse moment (iMOM) priors for the regression parameters $\boldsymbol{\beta}_k$, is given by

$$\pi_I(\boldsymbol{\beta}_k|\upsilon,\tau_2) = \frac{(\phi\tau_2)^{\upsilon n_k/2}}{\left(\Gamma\left(\frac{\upsilon}{2}\right)\right)^{n_k}}\prod_{j=1}^{n_k}|\beta_{j,k}|^{-(\upsilon+1)}\exp\left[-\phi\tau_2\sum_{j=1}^{n_k}\frac{1}{\beta_{j,k}^2}\right],$$

where $\tau_2$ is the common scale parameter and $\upsilon$ is the common shape parameter of the iMOM priors.

The peMOM prior, which is the product of individual exponential moment (eMOM) priors for the regression parameters $\boldsymbol{\beta}_k$, is given by

$$\pi_E(\boldsymbol{\beta}_k|\tau_3) = c\,(\phi\tau_3)^{-\frac{n_k}{2}}\exp\left\{-\sum_{j=1}^{n_k}\frac{\phi\tau_3}{\beta_{j,k}^{2r}}\right\}\prod_{j=1}^{n_k}\exp\left[-\frac{1}{2\phi\tau_3}\sum_{j=1}^{n_k}\beta_{j,k}^2\right],$$

where $\tau_3$ is the common scale parameter and $r$ is the common order of the eMOM priors and $c = (2\pi)^{-n_k/2}e^{n_k\sqrt{2}}$ is a constant.

For the model space, we use a beta-binomial prior [3], given by $\pi(k|\gamma) = \gamma^{n_k}(1-\gamma)^{p_s-n_k}$ where $\gamma \sim beta(1,1)$ distribution. Model selection in the Bayesian paradigm generally entails computation of the posterior probability for each possible model and selecting the model with the highest posterior probability.

## II. C. THE PROPOSED METHOD

The proposed iterative variable selection method adopts the recently proposed structured screen-and-select strategy [2]. Structured screening considers association between the outcome and the covariates as well as association among the covariates. For the AFT model, logarithm of the observed failure times are the outcomes. The proposed method is described as follows.

- In the first iteration ($iter = 1$), the association of all candidate covariates with the outcome is ascertained in terms of marginal utility (MU) [5]. For the $j$th candidate variable, the MU is the maximum partial likelihood for that variable given by

$$mu_j = \sum_{i=1}^{n}\left[-\frac{\delta_i}{2\sigma^2}\{\log(y_i) - (\mu + x_{j,ik}\beta_{j,k})\}^2 - (1-\delta_i)\,\Phi\left\{\frac{\log(y_i) - (\mu + x_{j,ik}\beta_{j,k})}{\sigma}\right\}\right].$$

We rank the variables according to their marginal utility and call the top $k_0$ variables having the largest absolute marginal utility the *leading variables*. Next, for a given threshold $r$, for each leading variable, we collect all candidate variables that have an absolute Pearson correlation coefficient value $r$ with that leading variable; these collections are called the *leading sets*. For each leading set, non-local prior-based model selection for the AFT survival model is performed according to the method in [4] using the non-local priors described above. The variables contained in the higher posterior probability model of each leading set are included in the final selection. The remaining variables of each leading set are excluded from further consideration.

- Subsequent iterations, $iter = 2, 3, ...$, proceed as before except for the evaluation of the association between candidate variables and the outcome which is done as follows. Suppose $X_{sel}^{(iter-1)}$ is the set of variables selected in iterations 1 through $i-1$, and $x_{i,sel}^{(iter-1)}$ is its $i$th row. In iteration $iter$, the association of the remaining candidate variables with the binary outcome is ascertained in terms of their conditional utilities (CU) in presence of the variables $X_{sel}^{(iter-1)}$ in the model [5]. For the $j$th candidate variable of iteration $iter$, the CU is given by

$$cu_j^{iter} = \sum_{i=1}^{n} \left[ -\frac{\delta_i}{2\sigma^2} \left\{ \log(y_i) - \left( \mu + x_{i,sel}^{(iter-1)} \boldsymbol{\beta}_{sel}^{(iter-1)} + x_{j,ik}^{iter} \beta_{j,k} \right) \right\}^2 \right. \\ \left. - (1-\delta_i) \Phi \left\{ \frac{\log(y_i) - \left( \mu + x_{i,sel}^{(iter-1)} \boldsymbol{\beta}_{sel}^{(iter-1)} + x_{j,ik}^{iter} \beta_{j,k} \right)}{\sigma} \right\} \right].$$

- Variables are selected through this structured screen-and-select strategy until we reach a desired number, $m$, of selected variables, or until the number of iterations that select no variables reaches a maximum allowed value, $maxno$.

The number of selected variables by the proposed method depends on four tuning parameters $k_0$, $r$, $m$ and $maxno$. In the following application, we set them using the heuristic guidelines provided in [2].

## III. APPLICATION

We compared the proposed method with few existing methods in a simulation study. We considered 1000 subjects and 10000 covariates. Without loss of generality, the first 6 covariates were chosen as true covariates having non-zero regression coefficients whereas all other covariates had zero coefficient. The non-zero coefficients were taken as (0.8, -0.9, 1.3, -1.4, 0.5, -0.53). The design matrix was randomly generated from N(0,1) distribution. Given the design matrix and the regression coefficients, we simulated survival times lying in between 0 to 20 from two different models—from the AFT model (using R package *imputeYn*) with 50% censored observations and from the Cox proportional hazards (PH) model (using R package *coxed*) with 30% censored observations for examining performance under model misspecification.

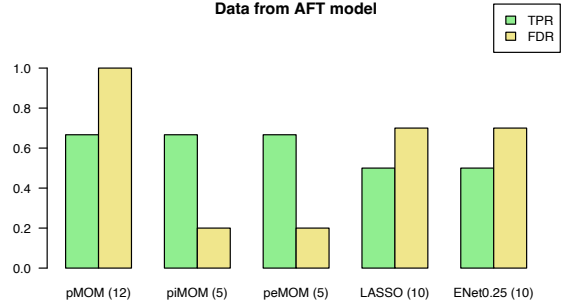

**Figure 1.** Performance comparison of various methods for data generated from the accelerated failure time (AFT) model. The number of selected variables by the different methods are given inside parentheses beside their names.

We perform variable selection in the simulated datasets using our proposed method, LASSO and Elastic net (α=0.25). For our proposed method, we used $\phi = 1$, $\tau_1 = 0.01$, $\tau_2 = 0.01$, and $\tau_3 = 0.01$ for data from the AFT model and $\phi = 1$, $\tau_1 = 0.192$, $\tau_2 = 0.25$, and $\tau_3 = 0.091$ for data from the Cox PH model following suggestions provided in [1] and [4]. The tuning parameters of the proposed method were set as $k_0 = 1$, $r = 0.2$, and $maxno = 3$. For LASSO and elastic net (α=0.25) analyses, we used the *AEnet.aft* function in the R package *AdapEnetClass* with default options. The performance of all the methods was compared in terms of true positive rate (TPR) and false discovery rate (FDR), and are shown in Figure 1 (for data from the AFT model) and Figure 2 (for data from the Cox PH model). Our proposed method with peMOM prior has shown the best performance in these datasets. The performance of the piMOM prior based analysis comes next.

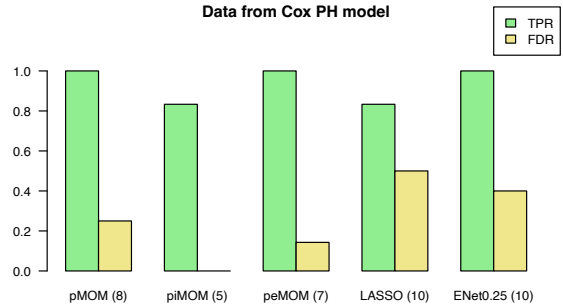

**Figure 2.** Performance comparison of various methods under model misspecification for data generated from the Cox proportional hazards (PH) model. The number of selected variables by the different methods are given inside parentheses beside their names.

## CONCLUSION

We have proposed an iterative variable selection method in accelerated failure time models for high-dimensional survival data based on the recently proposed structured screen-and-select framework and nonlocal priors. The proposed method has been implemented within the existing R package *GWASinlps*. In this work, we have set the scale parameters of the nonlocal priors and the tuning parameters of the proposed method heuristically. Future work will focus on developing more objective cross-validation or other data-dependent methods for setting these parameters.